\documentstyle[twocolumn,aps]{revtex}

\begin{document}
\draft
\title{\bf Ultrastable ${\rm CO}_{2}$ Laser Trapping of Lithium Fermions}
\author{K. M. O'Hara, S. R. Granade, M. E. Gehm, T. A. Savard, S.
Bali,\\ C. Freed$^{\dagger}$, and J. E. Thomas}
\address{Physics Department, Duke University,
Durham, North Carolina 27708-0305}
\date{March 8, 1999}
\wideabs{
\maketitle
\begin{abstract}
We demonstrate an ultrastable ${\rm CO}_{2}$ laser trap that
provides tight confinement of neutral atoms with negligible
optical scattering and minimal laser noise induced heating. Using
this method, fermionic $^{6}{\rm Li}$ atoms are stored in a 0.4 mK
deep well with a 1/e trap  lifetime of 300 seconds, consistent
with a background pressure of $10^{-11}$ Torr. To our knowledge,
this is the longest storage time ever achieved with an all-optical
trap, comparable to the best reported magnetic traps.
\end{abstract}

\pacs{PACS numbers: 32.80.Pj \\ Copyright 1999 by the American
Physical Society \\}
}

Off-resonance optical traps have been explored for many years
as an attractive means of tightly confining neutral
atoms~\cite{Ashkin}. Far off resonance optical traps (FORTs)
employ large detunings from resonance  to achieve
low optical heating rates and high density, as well as to
enable trapping of multiple atomic spin states in nearly identical
potentials~\cite{Heinzen,Phillips,Chu1,Chu2,Ketterle}.
For  ${\rm CO}_{2}$ laser traps~\cite{Knize}, the extremely large detuning
from resonance and the very low optical frequency lead to optical
scattering rates that are measured in photons per atom per {\em hour}. Hence,
optical heating is negligible.
Such traps are potentially important for development
of new standards and sensors based on spectroscopic methods, for
precision measurements such as determination of  electric dipole
moments in atoms~\cite{Fortson}, and for fundamental
studies of cold,  weakly interacting  atomic or molecular vapors.

However, all-optical atom traps have suffered from unexplained
heating mechanisms that limit the minimum attainable temperatures
and the maximum storage times in an ultrahigh vacuum~\cite{Chu1,Chu3,Adams}.
Recently, we have shown that
to achieve long storage times in all-optical traps that are not limited
by optical heating rates, heating arising from laser intensity noise
and beam pointing noise must be stringently controlled~\cite{Savard,Gehm}.
Properly designed ${\rm CO}_{2}$ lasers are powerful and extremely
stable in both frequency and intensity~\cite{Freed,Thomas},
resulting in laser-noise-induced heating times that  are measured in hours.
Hence,  in an ultrahigh vacuum (UHV) environment, where loss and heating
arising from background gas collisions are
minimized~\cite{Monroe,Bali},
extremely long storage times should be obtainable using ultrastable
${\rm CO}_{2}$ laser traps.

In this Letter, we report storage of  $^{6}{\rm Li}$ fermions
in an ultrastable ${\rm CO}_{2}$ laser trap. Trap 1/e lifetimes
of 300 seconds are obtained, consistent with
a background pressure of $10^{-11}$ Torr. This constitutes
the first experimental proof of
principle that extremely long storage times can be achieved in
all-optical traps. Since arbitrary hyperfine states
can be trapped, this system  will enable exploration
of s-wave scattering in a weakly interacting fermi gas.

The  well-depth for a focused ${\rm CO}_{2}$ laser trap
is determined by the induced dipole potential
$U=-\alpha_{g}\bar{{\cal E}^{2}}/2$, where $\alpha_{g}$
is, to a good approximation, the ground state {\em static}
 polarizability~\cite{Knize},
and $\bar{{\cal E}^{2}}$ is the time average of the square of the
laser field. In terms of the maximum laser intensity $I$ for the
gaussian ${\rm CO}_{2}$ laser beam,  the ground state
well-depth $U_{0}$ in Hz is
\begin{equation}
\frac{U_{0}}{h}({\rm Hz})=-\frac{2\pi}{hc}\,\alpha_{g}\,I .
\label{eq:depth}
\end{equation}
In our experiments,  a laser power of P=40 W typically is
obtained  in the trap region. A lens is used to focus the trap beam
to field a 1/e radius of $a_{f}=50\,\mu$m, yielding an
intensity of $I=2P/(\pi a_{f}^{2})\simeq 1.0\,{\rm MW}/{\rm cm}^{2}$.
For the I-P(20) line with $\lambda_{{\rm CO}_{2}}\simeq10.6\,\mu$m,
the Rayleigh length is $z_{0}=\pi a_{f}^{2}/\lambda_{{\rm CO}_{2}} =0.74$ mm.
Using the Li ground state polarizability of $\alpha_{g}=24.3\times
10^{-24}\,{\rm cm}^{3}$~\cite{polarize} yields a well
depth of $U_{0}/h=-8$ MHz, which is approximately 0.4 mK.
For this tight trap, the  $^{6}{\rm Li}$
radial oscillation frequency is 4.7 kHz and the axial frequency is 0.22 kHz.

For $^{6}{\rm Li}$ in a
 ${\rm CO}_{2}$ laser trap,  both the excited
 and the ground states are attracted to the well.
 The excited state static polarizability is $\alpha_{p}=18.9\times
 10^{-24}{\rm cm}^{3}$~\cite{polarize}, only 20\% less than that of
 the ground state. With a ground state well depth of 8 MHz,
 the frequency of the first resonance transition in the trap is shifted  by
 only 1.6 MHz at the center of the trap and thus does not significantly
 alter the  operation of the magneto-optical trap (MOT) from
 which the trap is loaded.

The optical  scattering rate $R_{s}$ in the  ${\rm CO}_{2}$ laser trap
 arises from Larmor scattering~\cite{Knize} and can be written as
 $R_{s}=\sigma_{S}\,I/(\hbar ck)$,
  where the Larmor scattering cross section $\sigma_{S}$ is
 \begin{equation}
 \sigma_{S}=\frac{8\pi}{3}\alpha_{g}^{2}k^{4} .
 \label{eq:xsection}
 \end{equation}
 Here, $k=2\pi /\lambda_{{\rm CO}_{2}}$.
 Using  $\alpha_{g}=24.3\times 10^{-24}\,{\rm cm}^{3}$ yields
 $\sigma_{S}=5.9\times 10^{-30}{\rm cm}^{2}$. At 1.0 MW/${\rm cm}^{2}$,
 the scattering rate for lithium is then $2.9\times 10^{-4}$/sec, corresponding
 to a scattering time of $\simeq 3400$ sec for one photon per atom.
 As a result, the recoil heating rate is negligible.

 Heating can arise from laser intensity noise and beam pointing
 fluctuations~\cite{Savard,Gehm}. For simplicity, we estimate the
  noise-induced heating rates for our trap using  a harmonic oscillator
  approximation  which is valid for atoms  near the bottom of the well.
 This provides only a rough estimate of the expected
 heating rates in the gaussian well, since the trap oscillation
 frequency decreases as the energy approaches the top of the well.
 A detailed discussion
 of noise-induced heating in  gaussian potential wells will be
  given in a future publication.  In the harmonic oscillator approximation,
 intensity noise causes parametric  heating and an exponential
 increase in the average energy for  each  direction of oscillation,
 $\langle\dot{E}\rangle =\Gamma \langle E \rangle$, where the rate
 constant in ${\rm sec}^{-1}$ is
 \begin{equation}
 \Gamma =\pi^{2}\nu^{2}\,S_{I}(2\nu ) .
 \label{eq:gamma}
 \end{equation}
 Here $\nu$ is a trap oscillation frequency and $S_{I}(2\nu )$ is the
 power spectrum of the fractional intensity noise in ${\rm
 fraction}^{2}/{\rm Hz}$.  For our ${\rm CO}_{2}$ laser,
 $S_{I}(9.4\,{\rm kHz})\leq 1.0\times10^{-13}$/Hz, where it is
 comparable to the detector noise. This is three orders of magnitude
 lower than that measured for an argon ion laser~\cite{Savard}.
 The corresponding heating time for  radial oscillation in our trap at
$\nu =4.7$ kHz is  $\Gamma^{-1}\geq 4.6\times 10^{4}$ sec.
 For the axial oscillation, $\nu= 220$ Hz,
 $S_{I}(440\,Hz)\simeq 1.1\times 10^{-11}$/Hz
and $\Gamma^{-1}\simeq 2\times 10^{5}$ sec.

 Fluctuations in the position of the trapping laser beam
 cause a constant heating rate  $\langle\dot{E}\rangle =\dot{Q}$, where
  \begin{equation}
 \dot{Q}=4\pi^{4}M\nu^{4}S_{x}(\nu ) .
 \label{eq:position}
 \end{equation}
 Here $M$ is the atom mass and $S_{x}$ is the position noise power
 spectrum in ${\rm cm}^{2}$/Hz at the trap focus.
 For $^{6}{\rm Li}$, one obtains  $\dot{Q}({\rm nK/s})=2.8\times
 10^{-4}\nu^{4}(Hz)\,S_{x}(\mu{\rm m}^{2}/{\rm Hz})$. Position
 noise only couples directly to the radial motion where $\nu\simeq 4.7$ kHz.
 For our laser, $S_{x}(4.7\,{\rm kHz})\leq 3.4\times10^{-10}\,\mu{\rm m}^{2}$/Hz,
 where the upper bound is determined by the noise floor for our
 detection method.  This yields $\dot{Q}\leq$ 46 nK/s. Hence,
 we expect the trap lifetime to be limited by
 the background pressure of our UHV system.

 The expected   number of trapped atoms $N_{T}$ can be estimated as follows.
 We take the  trapping potential to be approximately gaussian in
 three dimensions:
 \begin{equation}
 U(\vec{x})=-U_{0}\,\exp (-x^{2}/a^{2}-y^{2}/b^{2}-z^{2}/z_{o}^{2}) ,
 \label{eq:gaussian}
 \end{equation}
 where $a=b=a_{f}/\sqrt{2}$ is the intensity 1/e radius and $z_{o}$ is
 the Rayleigh length. Here, the lorentzian dependence
 of the trap beam intensity on the axial position $z$ is
 approximated  by a gaussian dependence on $z$.

 We assume that after a  sufficient loading time,
 atoms in the ${\rm CO}_{2}$ laser trap will come into thermal and diffusive
 equilibrium with the MOT atoms that serve as a reservoir~\cite{O'Hara2}.
 The density of states in the gaussian trap and the occupation number
 then determine  the number of trapped atoms, which takes the form
 \begin{equation}
 N_{T}=n\,V_{FORT}\,F[U_{0}/(k_{B}T)] .
 \label{eq:trapnum}
 \end{equation}
 Here the volume of the ${\rm CO}_{2}$ laser trap is defined as
 $V_{FORT}=a^{2}z_{o}\pi^{3/2}$. Hence,  $n\,V_{FORT}$ is the total number
 of atoms contained in  the volume of the FORT  at the MOT  density $n$.

 $F(q)$ determines the number of  trapped atoms
compared to the total number contained in the FORT volume
 at the MOT density. It is a function only of the ratio
 of the well depth to the MOT temperature,  $q\equiv U_{0}/(k_{B}T)$:
 \begin{equation}
 F(q)=\frac{q^{3/2}}{2}\int_{0}^{1}dx\,x^{2}\,g(x)\,\exp [q(1-x)] .
 \label{eq:enhance}
 \end{equation}
 Here $g(x)$ is the ratio of the density of states for
 a gaussian well to that of a three dimensional harmonic well:
 \begin{equation}
 g(x)=\frac{\beta^{3/2}(1-x)^{1/2}}{x^{2}}\frac{16}{\pi}
 \,\int_{0}^{1}du\,u^{2}\sqrt{\exp [\beta (1-u^{2})]-1} ,
 \label{eq:ratio}
 \end{equation}
 where $\beta\equiv -\ln (1-x)$.
 The variable $x=(E+U_{0})/U_{0}$
  is the energy of the atom relative to the bottom
 of the well in units of the well depth, where $-U_{0}\leq E\leq 0$,
 and $g(0)=1$.
 For  our MOT, the typical temperature is 1 mK,
 $n\simeq 10^{11}/{\rm cm}^{3}$,
 and $n\,V_{FORT}=5\times 10^{5}$ atoms.
 Using the well depth of $U_{0}=0.4$ mK in  Eq.~\ref{eq:trapnum} shows
 that $N_{T}$ is of the order of $6\times 10^{4}$ atoms.
 Much higher numbers are obtainable for a deeper well at lower temperature.

The experiments  employ a custom-built, stable ${\rm CO}_{2}$
laser. High-voltage power supplies,
rated at $10^{-6}$ fractional stability at full voltage, proper
electrode design, and negligible plasma noise enable highly stable current.
Heavy mechanical construction, along with thermally and acoustically shielded
invar rods, reduces vibration.
The laser produces  56 W in an excellent  ${\rm TEM}_{00}$ mode.

The ${\rm CO}_{2}$ laser beam is
expanded using a ZnSe telescope. It is focused through a double-sealed,
differentially-pumped, 5 cm diameter ZnSe  window into a
UHV system. The vacuum is maintained at $\simeq 10^{-11}$ Torr by a titanium
sublimation pump. The trap is at the focus
of a 19 cm focal length ZnSe lens.

The trap is continuously loaded from  a $^{6}{\rm Li}$ MOT employing a standard
$\sigma_{\pm}$ configuration~\cite{MOT} with
  three orthogonal pairs of counterpropagating, oppositely-polarized
  671 nm laser beams, each 2.5 cm in diameter and 8 mW.  Power is supplied by
 a  Coherent 699 dye laser that generates 700 mW. The MOT magnetic field
  gradient is 15 G/cm (7.5 G/cm) along the radial (axial) directions
  of  the trap.  The MOT is loaded from a multicoil Zeeman
  slower system~\cite{multicoil} that employs a differentially pumped
  recirculating oven   source~\cite{recirc}.
   Using a calibrated photomultiplier,
  the MOT is estimated to trap approximately $10^{8}$ $^{6}{\rm Li}$ atoms.
  The MOT volume is found to be  $\simeq 1\,{\rm mm}^{3}$. This yields
 a density of $10^{11}/{\rm cm}^{3}$,  consistent with that obtained for lithium
 in   other experiments~\cite{Shimizu,Anderson}. Using
 time-of-flight methods, we find typical MOT  temperatures of 1 mK.

 We initially align the ${\rm CO}_{2}$ laser
  trap with the MOT by using  split-image
 detection of the fluorescence at 671 nm to position the
  focusing ZnSe lens in the axial  direction.
  The focal point for the trapping beam is positioned in the center of the
 MOT, taking into account  the difference in the index of
 refraction of the optics at 671 nm and  10.6 $\mu$m.
 A 671 nm laser  beam is  aligned on top of the ${\rm CO}_{2}$
 laser beam to align the transverse position
 of the focal point in the MOT.  Since the Rayleigh length
 is short and the focus is tight, this method does not reliably locate
 the actual focus of the ${\rm CO}_{2}$ beam. Hence, a spectroscopic
 diagnostic based on the  light shift induced
 by the ${\rm CO}_{2}$ laser is employed for final alignment of the
 trapping beam.

 While the near equality of the Li excited and ground state polarizabilities
 is ideal for continuous loading from the MOT, it
 makes locating the ${\rm CO}_{2}$ laser focus in the MOT by light
 shift methods quite  difficult.
 To circumvent this problem, a dye laser at  610 nm
 is used to excite the 2p-3d transition  for  diagnostics.
 At the 10.6 $\mu$m ${\rm CO}_{2}$ laser  wavelength, we estimate that
 the 3d state has a scalar polarizablity of approximately
 $700\times 10^{-24}{\rm cm}^{3}$~\cite{Litransitions},
 nearly 30 times that of the 2s or
 2p state. In the focus of the ${\rm CO}_{2}$ laser, the corresponding
 light  shift is $\simeq -300$ MHz. Chopping the ${\rm CO}_{2}$ laser
 beam at 2 kHz  and using lock-in detection of fluorescence
 at 610 nm yields a two-peaked light shift spectrum.
 This two-peaked structure arises because the lock-in  yields
 the difference between signals with the ${\rm CO}_{2}$ laser blocked
 and unblocked. At the ideal focusing lens position, the  amplitude
 and the frequency separation of these peaks are maximized. Optical
alignment  remains unchanged for months after this procedure.

Measurement of  the trapped atom number versus time
is accomplished  by monitoring the fluorescence at 671 nm induced by a pulsed,
retroreflected probe/repumper beam (1 mW, 2 mm diameter). The probe
is double-blinded by acousto-optic (A/O) modulators to minimize trap loss arising
from probe light leakage. The loading sequence is as follows:
First, the ${\rm CO}_{2}$ laser trap is continuously loaded
from the MOT  for 10 seconds. This provides adequate time for
the MOT to load from the Zeeman slower.
Then the MOT  repumping beam is turned off, so that atoms in the upper
$F=3/2$ hyperfine state are optically pumped into the lower $F=1/2,M=\pm 1/2$
states.  After 25 $\mu$sec, the optical MOT beams
are turned off using A/O modulators, and a mechanical shutter in front
of the dye laser  is closed within 1 ms to eliminate all MOT light at 671 nm.
The MOT gradient magnets are turned off within 0.2 ms.
After a predetermined time interval between  0 and 600 sec,
the probe beam is pulsed to yield a fluorescence
signal proportional to the  number of trapped atoms. The detection system
is calibrated and the solid angle is estimated to determine the atom
number. Typical trapped atom numbers
measured in our initial experiments are $\simeq 2.3\times 10^{4}$.
This corresponds to the predictions of  Eq.~\ref{eq:trapnum} for a
well depth of 0.25 mK. Since we expect the potential of the
 MOT gradient magnet
to lower the effective well depth from 0.4 mK by $\simeq 0.15$ mK
during loading, the measured trap number is consistent with our predictions.

Fig.~\ref{fig:1} shows the decay of the trapped atom number on
a time scale of 0-600 seconds. Each data point is the mean obtained
from four separate measurement sequences through the complete decay curve. The
error bars are the standard deviation from the mean.
Atoms  in the $F=1/2$ state exhibit a single exponential
decay  with a time constant of 297 sec, clearly demonstrating
the potential of this system for measurements on a long time scale.

We have observed that an initial 10-15\% decrease in the
signal can occur during the first second. This may arise from
inelastic collisions between atoms in the $F=1/2$ state
with atoms that are not optically pumped out of the upper $F=3/2$ state.
During optical pumping, fluorescence from the F=3/2 state
decays in $\simeq 5\,\mu$sec to a $\simeq 5$\%  level
which persists for a few milliseconds, consistent with a residual $F=3/2$
population.

The lifetime of atoms in the $F=1/2$ state can be limited by processes
that cause heating or direct loss.
If we attribute the trap lifetime entirely to residual heating, the
heating rate from all sources
would be at most $400\mu{\rm K}/300 {\rm sec}\simeq 1 \mu$K/sec,
which is quite small. However, if the  loss were due to heating,
one would expect a multimodal decay curve,
analogous to that predicted in Ref.~\cite{Gehm}. Instead, we observe
a single exponential decay as expected for
direct loss mechanisms, such as collisions with background gas atoms
or optical  pumping  by background light  at 671 nm
(into the unstable $F=3/2$ state).
If we assume that the lifetime is background gas limited and that Li
is the dominant constituent, the measured lifetime of 297 sec is
consistent with a pressure of $\simeq 10^{-11}$ Torr.

The long lifetime of the  $F=1/2$ state is  expected, based on
the prediction of a negligible s-wave elastic scattering length ($<<1$ Bohr)
at zero magnetic field~\cite{Stoof}. Hence, spontaneous evaporation
should not occur. We have  made a preliminary measurement
of trap loss arising from
inelastic collisions when the $F=3/2$ state is occupied. This is accomplished
by omitting the optical pumping  step in the loading
sequence described above. The trap is found to decay with a 1/e time
$<1$ sec when $2.3\times 10^{4}$ atoms are loaded (density $\simeq
10^{9}/{\rm cm}^{3}$). A detailed  study
of elastic and inelastic  collisions at low magnetic field is in progress.

In conclusion, we have demonstrated a 300 sec 1/e
lifetime for lithium fermions in an ultrastable ${\rm CO}_{2}$ laser trap
with a well depth of 0.4 mK.
By using an improved aspherical lens system,
an increase in trap depth to 1 mK is achievable. Further,
 Eq.~\ref{eq:trapnum} shows that, if the  MOT
temperature is reduced to 0.25 mK,
more than $10^{6}$ atoms can be trapped in a
1 mK deep well. Since the ground and excited
state trapping potentials are nearly identical, exploration
of optical cooling schemes may be particularly fruitful in this system.
Currently, we are exploring $^{6}{\rm Li}$ as a fundamental example
of a cold, weakly-interacting fermi gas. By trapping multiple
hyperfine states, it will be possible to study both
elastic and inelastic collisions between fermions. The combination
of long storage times and tight confinement obtainable
with the ${\rm CO}_{2}$ laser trap, as well as the anomalously large
scattering lengths for $^{6}{\rm Li}$~\cite{Hulet2,BCS}, make
this system an excellent candidate for evaporative cooling
and potential observation of a Bardeen-Cooper-Schrieffer transition.
Further, this system
is well suited for exploring novel wave optics of
atoms and molecules, such as  coherent changes of statistics by
transitions between free fermionic atoms and bosonic molecules,
analogous to free to bound transitions  for
bosonic atoms~\cite{Heinzen2}.

We thank Dr. R. Hulet for stimulating conversations regarding this work.
We are indebted to Dr. C. Primmerman and Dr. R. Heinrichs
of MIT Lincoln Laboratory for the loan of two
stable high voltage power supplies and to Dr. K. Evenson of NIST,
Boulder for suggestions regarding the laser  design.
This research has been supported
by the Army Research Office and the National Science Foundation.

$^{\dagger}$Permanent Address, Department of Electrical Engineering
and Computer Science, MIT, Cambridge, MA 02139.

\begin{figure}
\caption{Trapped number of atoms  versus time for an ultrastable
${\rm CO}_{2}$ laser trap. The solid line is  a single
exponential fit, $N(t)=A\,\exp(-t/\tau)$, and gives $\tau=297$ sec.
We believe that a small fraction of atoms are lost at short times
$\leq 1$ sec, (see insert, 0-10 sec) from collisions with atoms that
remain in the $F=3/2$ state after optical pumping. Hence, the first two points
at 0.1 and 0.3 sec are neglected in the fit.
The trap lifetime for the remaining $F=1/2$ atoms is 297 seconds,
to our knowledge the longest ever obtained with an all-optical trap.}
\label{fig:1}
\end{figure}


\begin{references}
\bibitem{O'Hara2}A model of the spatial loading dynamics of a MOT-loaded
gaussian laser beam trap will be discussed elsewhere.
\bibitem{recirc}The design of the recirculating oven source was
suggested to us by R. Hulet, Rice University.
\end{references}
\end{document}